\documentclass[
superscriptaddress,amsmath,amssymb,aps,twocolumn,
notitlepage,
]{revtex4-2}

\usepackage{appendix}
\usepackage{graphicx}
\usepackage{dcolumn}
\usepackage{bm}
\usepackage{lipsum}
\usepackage[dvipsnames]{xcolor}
\usepackage[T1]{fontenc}
 \fontfamily{cmr}

\usepackage{float}
\usepackage{hyperref}
\usepackage{titlesec}
\usepackage{lipsum}
\usepackage{tikz-cd}
\usepackage{todonotes}
 
\tikzcdset{every label/.append style = {font = \small}}

\titleformat*{\section}{\large\bfseries}
\usepackage[normalem]{ulem}
\hypersetup{
    colorlinks=true,
    linkcolor=blue,
    citecolor=magenta,      
    urlcolor=cyan,
    }

\preprint{APS/123-QED}

\begin{document}

\title{An epsilon-near-zero-based nonlinear platform for ultrafast re-writable holography}

\author{M. Zahirul Alam}
\email{zahirul.alam@gmail.com}
\affiliation{Department of Physics, University of Ottawa, 25 Templeton Street, Ottawa, ON, K1N 6N5, Canada}

\author{Robert Fickler}
\email{robert.fickler@tuni.fi}
\affiliation{Tampere University, Photonics Laboratory, Physics Unit, Tampere, FI-33014, Finland}

\author{Yiyu Zhou}
\affiliation{Institute of Optics, University of
Rochester, Rochester, NY 14627, USA}
\affiliation{currently with the Department of Electrical Engineering, Yale University, New Haven, CT 06510, USA}

\author{Enno Giese}
\affiliation{Technische Universität Darmstadt, Fachbereich Physik, Institut für Angewandte Physik, Schlo{\ss}gartenstr. 7, 64289 Darmstadt, Germany}

\author{Jeremy Upham}
\affiliation{Department of Physics, University of Ottawa, 25 Templeton Street, Ottawa, ON, K1N 6N5, Canada}

\author{Robert W. Boyd}
\affiliation{Department of Physics, University of Ottawa, 25 Templeton Street, Ottawa, ON, K1N 6N5, Canada}
\affiliation{Institute of Optics, University of
Rochester, Rochester, NY 14627, USA}

\begin{abstract}

We re-examine real-time holography for all-optical structuring of light and optical computation using a contemporary material: a subwavelength-thick, spatially unstructured film of indium tin oxide (ITO). When excited by spatially structured light at epsilon-near-zero frequencies, the film acts as an efficient and reconfigurable diffractive optical platform for all-optical modulation of light such as spatial structuring and optical computations. We demonstrate a few percent of absolute diffraction efficiency over greater than 300\;nm bandwidth around telecom wavelengths using a film four orders of magnitude thinner than and up to six orders of magnitude faster than standard holographic materials. Our findings highlight the potential of using epsilon-near-zero-based nanostructures for efficient modulation of spatially structured light and rapid prototyping without complex nanofabrication processes.

\end{abstract}

\maketitle

\noindent
\textbf{Introduction}\\
The unmatched speed and parallelism of light offers tremendous opportunities to address the energy-density, heat, and bandwidth constraints of electrical interconnects in a computing platform \cite{miller2017attojoule}. The interest in using optical systems for information processing is not new, as even a simple lens can passively perform parallel Fourier transforms \cite{goodman2005fourieroptics}.
Optical processing technologies have been widely investigated throughout the late twentieth century \cite{ragnarsson1970a_new_holographic,ambs2010optical, goodman2020short}, such as the use of holographic techniques and nonlinear wave mixing in Fourier space for convolution and correlation operations \cite{goodman1971introduction, stroke1972optical, gabor1971holography, yariv1978four}. Recent advances in metamaterials \cite{Zheludev2012} and metasurfaces \cite{Chen2016, kuznetsov2024roadmap}, have led researchers to leverage wave physics for novel approaches to information processing \cite{silva2014performing, kulce2021all, gigan2022imaging, wang2023image, lopez2023self, zhou2024optical} for applications in hardware acceleration\cite{solli2015analog}
and machine learning \cite{LeCun2015,Lin2017,wetzstein2020inference,abou2025programmable}. These demonstrations of optical processing using diffractive optics and metasurfaces rely largely on nanofabricated or 3D-printed static structures \cite{hu2024diffractive}.
Such structures are written once and then used as a fixed layer for image processing or classification tasks with limited tunability (on/off), e.g., through temperature \cite{cotrufo2024reconfigurable} or mechanical deformation \cite{zhang2021reconfigurable}. These demonstrations highlight the massive parallelization potential of free-space diffractive optical computing structures and metasurfaces by exploiting spatial modes, wavelengths, time, polarizations, spins, and the orbital angular momentum of light. However, one of the challenges in this aspect is to implement full dynamic reconfigurability and on-demand tunability of diffractive optical surfaces \cite{hu2024diffractive, kuznetsov2024roadmap}. 

In this work, we implement a real-time holographic protocol by exploiting the properties of topologically structured light fields \cite{shen2019optical,yao2011orbital} and the sub-picosecond, unity-order nonlinear optical response of a transparent conductive oxide at its epsilon near-zero (ENZ) frequencies \cite{Alam2016, reshef2019nonlinear,kinsey2019near}. We experimentally demonstrate, using an unpatterned ENZ thin film, a rapidly reconfigurable computing surface capable of ultrafast copying of spatial modes of light, broad-band operation over more than 300\,nm, and all-optical first-order differentiation.

Traditional real-time holographic materials, such as photo-refractive crystals \cite{marder1997design, Hesselink2004} and photo-polymers \cite{ blanche2010holographic, volodin1995highly, Ashley2001}, have been extensively studied for all-optical computation. However, these materials suffer from slow refresh rates (milliseconds to seconds) \cite{blanche2010holographic}, low index contrast, are typically several millimeters thick, and sometimes require the read and object beams to be of widely different frequencies -- e.\,g. deep-UV for writing and visible for reading \cite{marder1997design, tay_updatable_2008,Barbastathis2019}. In contrast to traditional holographic materials, when indium tin oxide (ITO) -- a degenerately doped transparent conducting oxide widely used in touch screens and photovoltaics --  is excited at its epsilon-near-zero frequencies, it exhibits superior performance: ITO can exhibit unity-order index variations for high-contrast diffraction patterns with sub-picosecond refresh times, and allows for operation using infrared light in both frequency-degenerate and frequency-non-degenerate configurations.  These findings suggest that ITO-based surfaces could enable a wide range of applications, including rapid prototyping without complex nanofabrication processes \cite{Ozcan2016}.

\begin{figure}
\centering
\includegraphics[width=1\columnwidth]{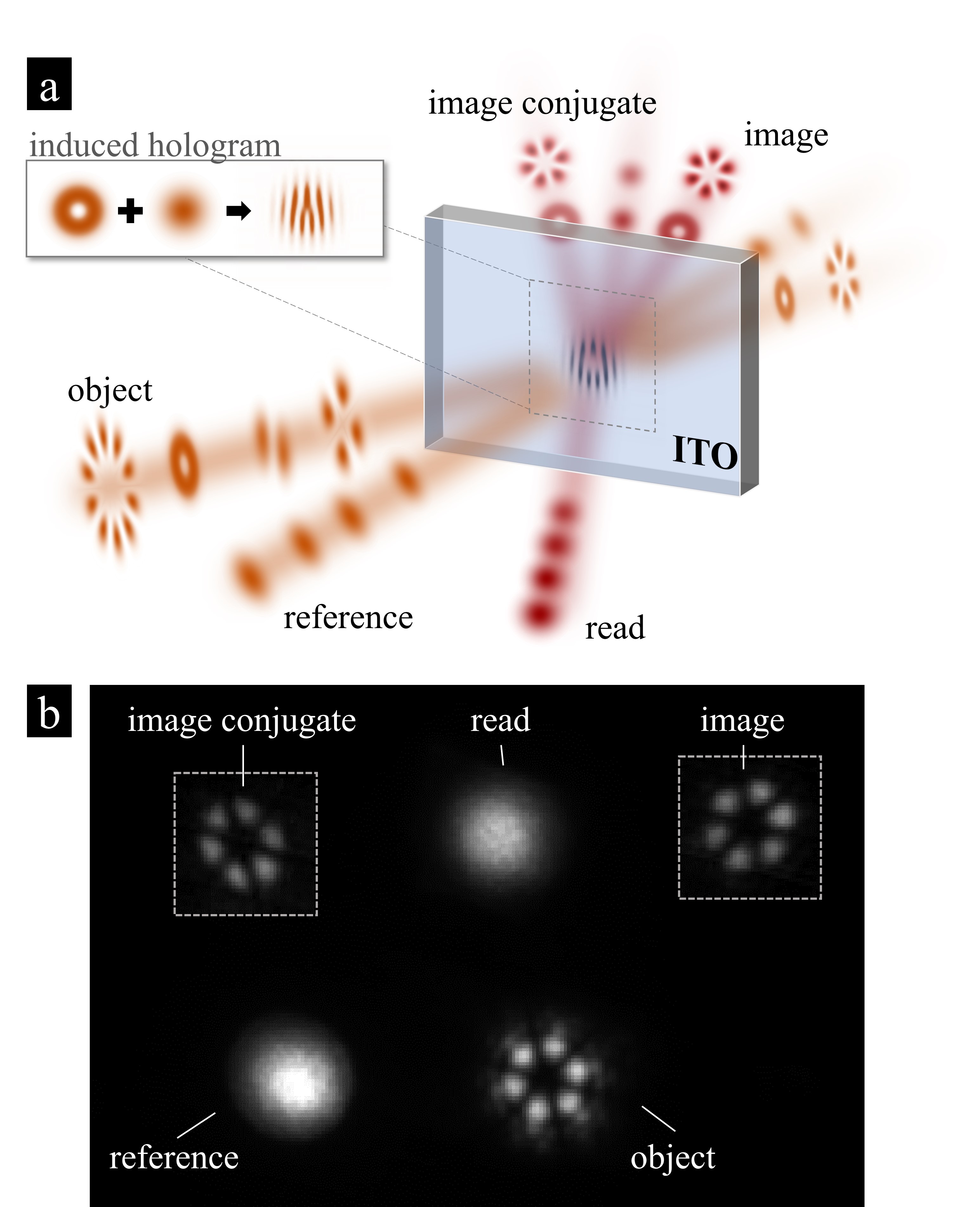}
\caption{\label{fig.experimental_config}
Experimental configuration.
(a) Interference of the object and the reference beams (orange) imprint a spatially varying hologram (a transient metastructure) onto the ITO film through an intensity-dependent change of refractive index (the example shown in the inset is for an object beam carrying one unit of OAM, that is, $\ell = 1$).
A read beam (red) of a distinct wavelength diffracts off the hologram.
The diffracted image (conjugate image) beam carries the same (complex conjugate) spatial structure as the original object beam.
(b) A camera image taken of the beams exiting the ITO film shows all relevant beams. 
For clarity, the two image beams were brightened by a factor of 10 (adjusted area indicated by dashed line). See Fig. S2 for the original unprocessed image.
}
\end{figure}

\noindent
\textbf{Transferring information between optical pulses}\\
The general scheme of our experiment is as follows. A Gaussian reference beam is brought to interfere with a frequency-degenerate object beam carrying an arbitrary image in the Fourier plane. All beams are $\sim$120~fs laser pulses and have a beam diameter of around 50-90~\textmu m at the plane of the ITO depending on the modal characteristics. A 310-nm-thick, transparent ITO film on a glass substrate is placed on the same Fourier plane. Due to the intensity-dependent response of the ITO's refractive index, the interference pattern induces a transient, spatially-varying, intensity-dependent refractive index distribution with high contrast between the unchanged (at the dark fringes) and nonlinearly changed index (at the bright fringes) \cite{Alam2016}. The induced transient metastructure, i.\,e., the hologram encoded using this technique carries the full spatial amplitude and phase information of the object beam. Furthermore, the encoding time of the interference pattern in the ITO thin film is less than 200 femtoseconds and it completely returns to the initial, uniform index within 1 picosecond. Consequently, a thin layer of ITO acts as an ultrafast, re-programmable diffractive metasurface -- programmed by the spatial patterns of the interfering structured light fields. Here we demonstrate such reconfigurability through two instructive applications, namely transduction of a spatial phase and intensity pattern to a read beam of varying wavelengths and edge-detection of an image using first-order differentiation.

In Fig.~\ref{fig.experimental_config}(a) we display a schematic representation of the holographic recording and image transduction processes using the induced hologram (shown in the inset) formed by the object and reference beams of wavelengths near the zero-permittivity wavelength of ITO, i.\,e. 1260\,nm. To read out the induced hologram ,we illuminate it with a read beam of the same wavelength but low power to operate in the linear regime with negligible influence on the induced hologram. The resulting linear diffraction of the read beam, results in two distinct copies of the initial object: the image and its conjugate counterpart. The image represents a reconstruction of the original object, while the counterpart is a complex conjugate image of the object. In practice, we placed a white screen in the far-field after the ITO and used an infrared camera to image the screen (see Fig.~S1 in the supplemental document for more details on the setup). Figure~\ref{fig.experimental_config}(b) is an example recording that captures all five beams, demonstrating that the intensity pattern of the object beam has been copied into the diffracted read beams (i.\,e. the image beam).

\begin{figure}
\centering
\includegraphics[width=\columnwidth]{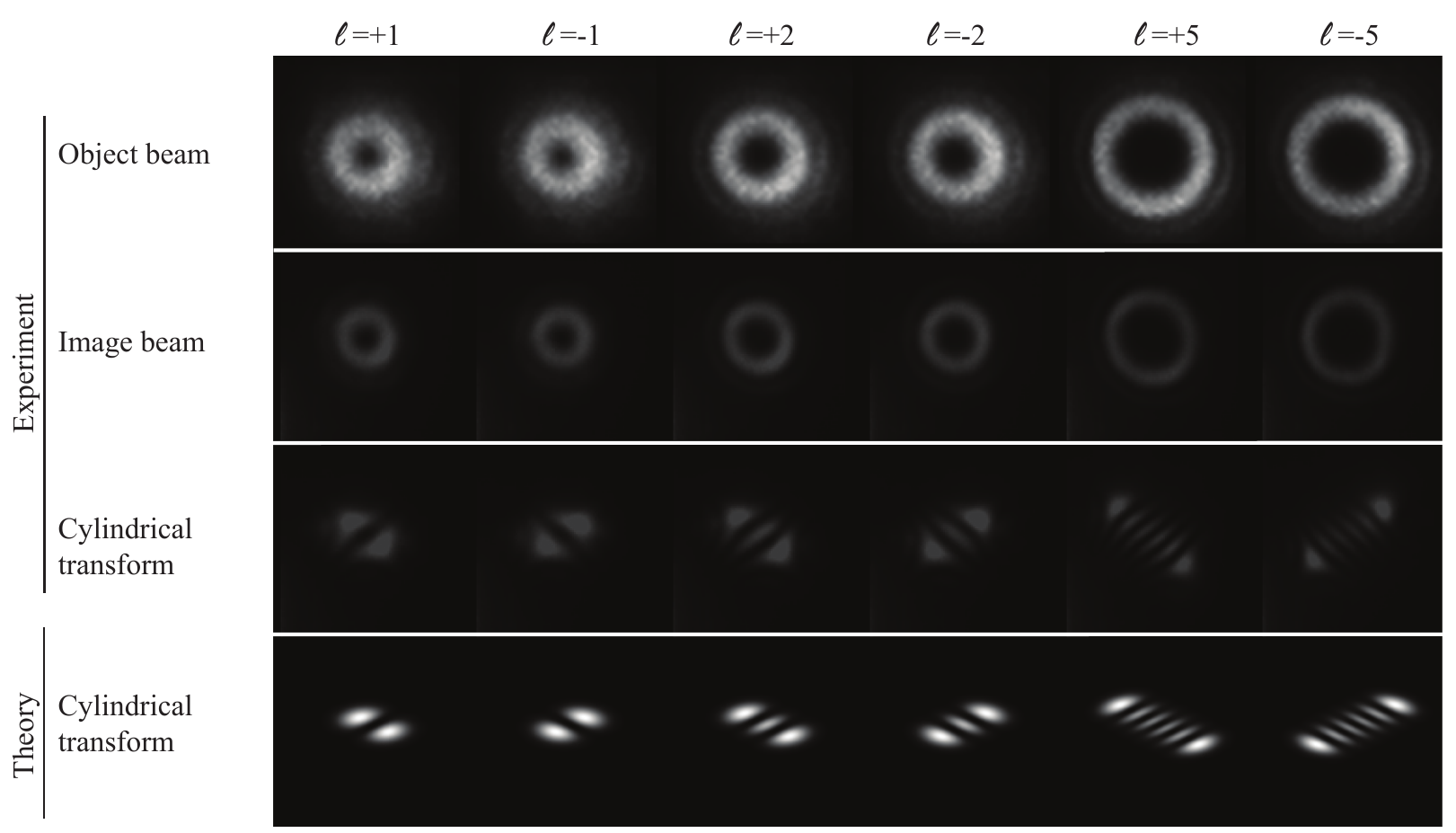}
\caption{
Transfer of spatial amplitude and phase information.
The induced holograms are sensitive to the spatial amplitude and phase of the object beams. To confirm the transfer of phase information, we prepare object beams with different signs and values of OAM ($\ell = \pm1,2,5$) and show the spatial structure of the resulting image beams. By performing cylindrical transformations and comparing the spatial structure and orientation of the transformed image beams with theory, we confirm the transfer of spatial phase information.}
\label{fig.transf_amplitude_phase}
\end{figure}

In addition to the intensity, we investigate the transfer of the spatial phase information to the read beam. We encode different orbital angular momentum (OAM)-carrying structures onto the object beam by shaping the light's transverse phase using a spatial light modulator. OAM-carrying light fields have an azimuthal phase gradient about the optical axis from 0 to 2$\pi\ell$, where $\ell$ corresponds to its integer OAM value \cite{allen1992orbital}. The twisted phase structure leads to a phase singularity and thus a vanishing intensity along the optical axis, resulting in a doughnut-like intensity pattern with a radius that increases with the OAM value (Fig.~\ref{fig.transf_amplitude_phase}, first row for different experimental recordings of the object beam). When detecting the resulting intensity patterns of the image beams (Fig.~\ref{fig.transf_amplitude_phase}, second row), we find that their intensity shapes match the corresponding original object beams, i.\,e. they show the same doughnut-shaped intensity structure that increases in radius with $\ell$. We verify the correct spatial phase distribution of the image beams by performing an astigmatic mode conversion using a cylindrical lens \cite{beijersbergen1993astigmatic}. The mode conversion process converts OAM beams into Hermite-Gauss (HG) modes that have a number of intensity lobes equal to $|\ell|+1$, arranged in a line oriented according to the sign of $\ell$ \cite{vaity2012topological}. Comparing the intensity patterns obtained after this cylindrical transform (Fig.~\ref{fig.transf_amplitude_phase}, third row) with the theoretical expectation (Fig.~\ref{fig.transf_amplitude_phase}, fourth row), we find very good agreement, verifying that we have transferred the amplitude and phase information of the light field from the object beam to the image beam.

\begin{figure*}
\centering
\includegraphics[width=2\columnwidth]{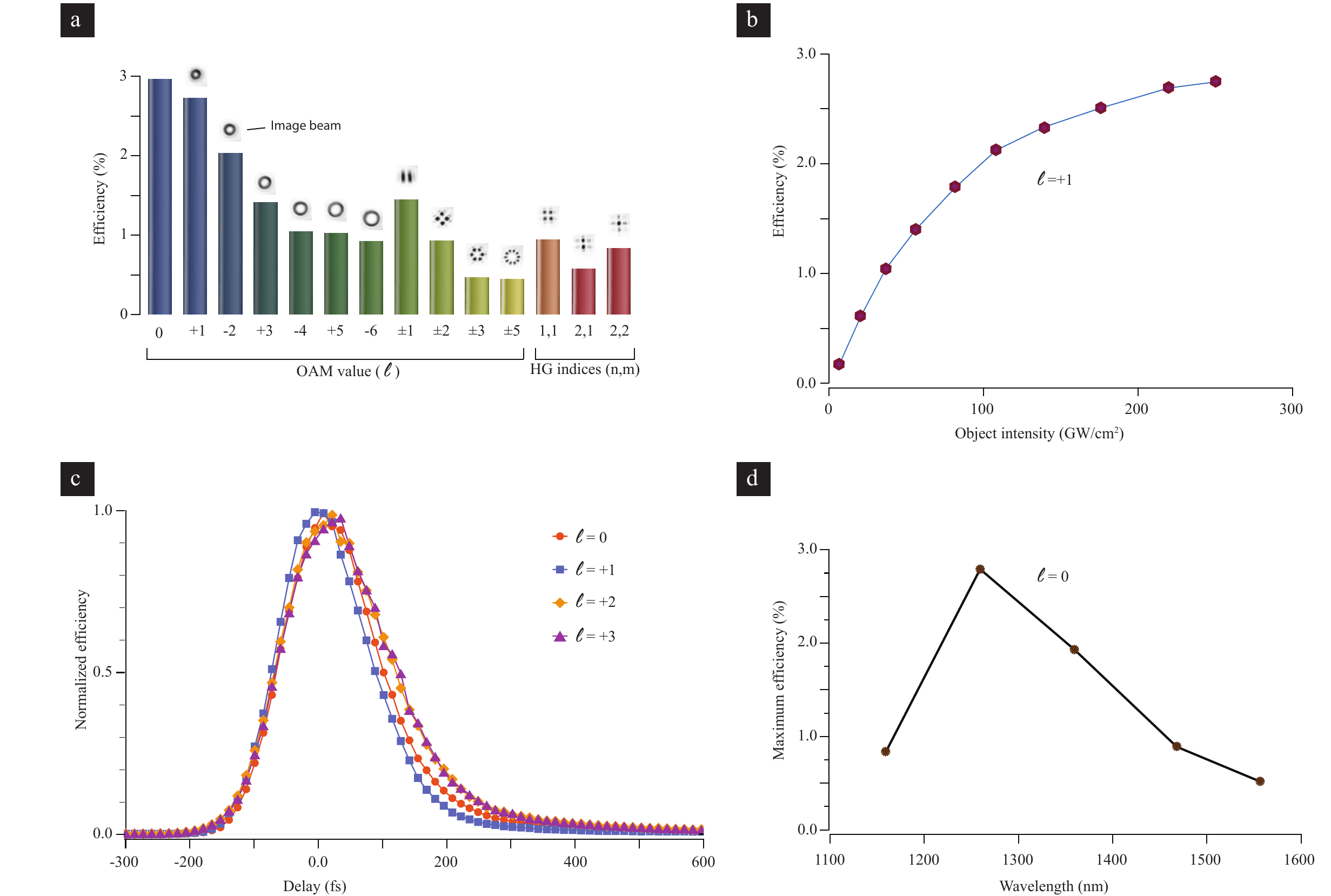}
\caption{
\label{fig.efficiency}
Efficiency of the information transfer process.
(a) The maximum efficiencies of various OAM modes, their superpositions, and a few Hermite-Gaussian modes.
(b) Dependence of information transfer efficiency on object beam intensity with $\ell=+1$. We set the intensity of the reference beam to be $\rm{300\,GW/cm^2}$ and measure the diffraction efficiency of the image beam as a function of the object beam intensity at the ITO layer.
(c) Temporal duration of the induced transient holographic structure. We vary the delay between the induced holographic structures (formed by the reference and object beams) and the read beam while recording the diffraction efficiency of the generated image beam for various spatial modes. We find that the hologram is only induced for the duration of a few hundred femtoseconds, suggesting a refresh rate of up to $\rm{1\,THz}$. 
(d) Tolerance of read beam wavelength. While the object and reference beams were set to 1260\,nm, we varied the read beam wavelength, showing significant image diffraction efficiency across more than 300\,nm.}
\end{figure*}

\noindent
\textbf{Quantifying transfer efficiency}\\
These initial experiments demonstrate the potential of ITO for applications in real-time holography. Next, we investigate the efficiency of the transfer process by measuring the diffracted power carried by the image beam. We measure this efficiency for several different object beams, including OAM modes, their superpositions, and a few low-order Hermite-Gauss modes. For these measurements, we set the power of the reference and object beams to saturate the nonlinear response of ITO (both at $\rm{300\,GW/cm^2}$, which is well below the damage threshold of roughly $\rm{2\,TW/cm^2}$), and also ensure that time delays between object, reference, and read beams are optimal. We then determine the efficiency by measuring the ratio between input read beam and resulting image beam through power meter measurements. The efficiencies are summarized in Fig.~\ref{fig.efficiency}(a), showing that the efficiency is maximum ($\approx$3\,\%,) for a Gaussian object beam ($\ell = 0$) and decreases with an increase in the radius ($|\ell| > 0$) of the object beam. Because the fringe visibility of the induced metastructure depends on the modal overlap of the object and reference beams and the diffraction efficiency of the read beam depends on its overlap with the metastructure, the maximum efficiency is obtained when all three incident beams have the same mode shapes. In our implementation these conditions are achieved for an $\ell = 0$ object beam, since then all three beams have Gaussian shapes. As a consequence of the reduced modal overlap between the three beams, object beams carrying higher-order OAM values ($|\ell| > 0$) or HG beams show lower efficiencies. While even the lowest efficiency of a few percent shown here is as good as holograms made from photo-refractive compounds \cite{mecher2002near}, we also note that the efficiency can be greatly increased through appropriate nano-structuring \cite{alam2018large}.

To gain insight into the intensity-dependent formation property of our holograms, we measure the efficiency of an $\ell=+1$ object beam by varying the power of the object beam while keeping the power of the reference Gaussian beam fixed at $\rm{300\,GW/cm^2}$ (Fig.~\ref{fig.efficiency}b). We find that efficiency increases as a function of the intensity of the object beam reaching a near-saturation value of approximately 3\,\%. The contrast of the induced hologram, and consequently the efficiency of the information transduction depends not only on the beam parameters but also on the intrinsic linear and the nonlinear properties of the ITO layer and the relative vectorial orientation of the interacting beams with respect to the ITO film. In general, an interference pattern of an object and the reference beam may induce three different but spatially overlapped patterns: 1) a phase grating due to modulation of the real part of the index; 2) an absorption grating due to modulation of the absorption (i.\,e. the imaginary part of the index); and 3) an amplitude grating due to strong modulation of the Fresnel reflection coefficients. A strong modification of the Fresnel reflection is expected, since the real part of the index of ITO can change from a value of 0.4 to nearly 1.2 \cite{Alam2016} resulting in strong angle-dependent linear Fresnel reflectance and a large nonlinear modulation of the reflectivity. Consequently, the measured efficiencies for various object beams are impacted by the  modulation of the complex optical constants of the ITO film and, thus, limited by the intrinsic saturation mechanisms of ITO's nonlinear response. For comparison, we note that the maximum diffraction efficiency of phase-only sinusoidal gratings operating in the Raman-Nath diffraction regime (thin hologram) is 33.9\,\% \cite{magnusson1978diffraction}. In our case the large modulation of the real part of the index ($\approx0.7$) \cite{Alam2016} can lead to a theoretical maximum diffraction efficiency of 22\,\%. However, these high values of the maximum diffraction efficiency are drastically reduced as a result of the superimposed amplitude and absorption gratings over the phase gratings.

Having established experimentally that an ITO thin film can be used to transfer information between beams of light, we consider how its ultrafast nonlinear optical response~\cite{Alam2016} can also offer the on-the-fly reconfigurability and rewritability using structured light fields. Figure~3\,c) demonstrates this ultrafast behaviour by the recorded efficiency of the holographic reconstruction while adjusting the delay between the formation of the hologram (i.\,e. the simultaneous arrival of the object and reference beams) and the read beam. We find that the non-zero efficiency sustains over a few hundred femtoseconds of delay before the index distribution returns to the unstructured initial state. This effect implies that ITO could accommodate refresh rates up to $\rm{1\,THz}$, that is, up to nine orders of magnitude higher compared to what can be obtained using any traditional photo-refractive material or polymers~\cite{tsutsumi2016molecular}. 
While single light fields rarely reach such high rate of modulations, the achievable refresh rates will enable various applications such as its simultaneous use for multiple object beams with minimal cross-talk, ultrafast temporal filtering, or its potential utilization in fields such as ultrafast transient holographic microscopy \cite{liebel2021ultrafast}.

Although the light-induced changes to the refractive index of ITO are the strongest near the zero-permittivity wavelength, we are not limited to using read beams at the same wavelength as the reference and object beams. Measuring the maximum diffraction efficiency ($\ell=0$) as a function of the read beam wavelength while keeping the reference and object beam fixed at 1260\,nm, we find that a maximum absolute diffraction efficiency of 1\,\% or more can be obtained in a wavelength range of more than 300\,nm (Fig.~\ref{fig.efficiency}d). We note that for wavelengths larger than the zero-permittivity wavelength, ITO becomes metal-like. Consequently, for wavelength larger than 1260\,nm, the Fresnel reflectivity and absorption increases (i.\,e. the net transmission decreases), while the phase grating becomes weaker. We also note that it is not imperative that the reference and object beam wavelengths be kept at precisely 1260\,nm. As long as they are within the above-mentioned wavelength range, the resulting large contrast between the base index and nonlinearly modulated index enable a reasonable diffraction efficiency of the read beam into the image beams. Nevertheless, the large operating bandwidth suggests the suitability of ITO information transfer from one color of light to another for applications in wavelength-division-multiplexing schemes.

\begin{figure}
\centering
\includegraphics[width=0.8\columnwidth]{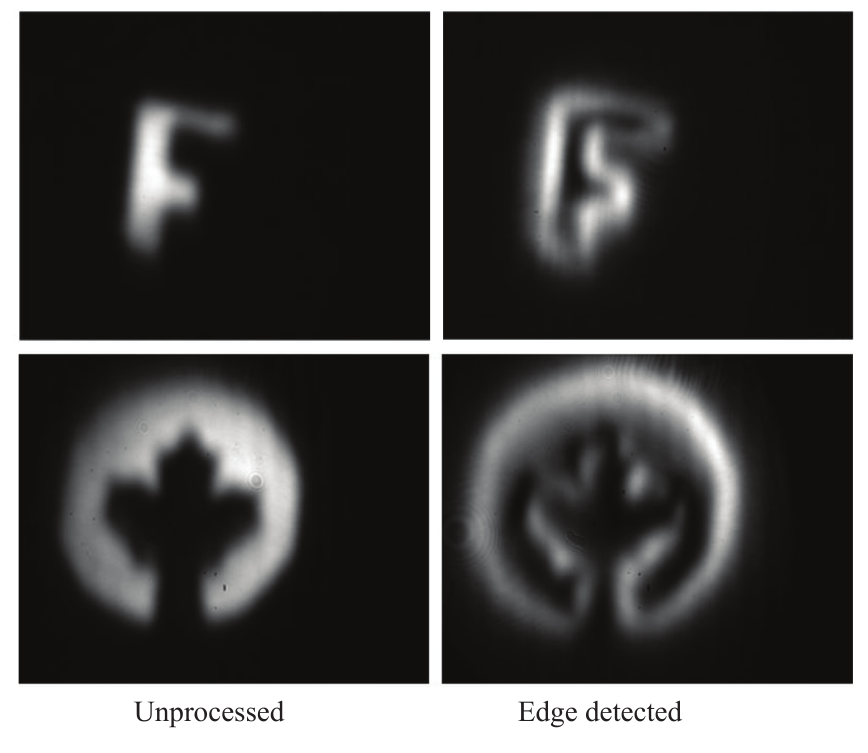}
\caption{
\label{fig.2d-differentiation}
Holographic 2D spatial first-order differentiation for edge detection.
When the object beam carries a topological charge ($\ell=+1$), the two-dimensional spatial differentiation of the image-carrying read beam is generated in the diffracted image beam. 
}
\end{figure}

\noindent
\textbf{Demonstrating all-optical operations}\\
The strong and ultrafast response of ITO not only enables holography far beyond video rates but also underlines that ITO is an ideal candidate for all-optical analog computation using transversely structured light fields. The general idea would be to use the read beam as an input to a computation device whose operation is configured by the transformation kernel -- the complex scattering properties of the metastructure -- induced by custom-tailored object and reference beams. The image beam is the output of the operation. To demonstrate the potential of our approach for such computations, we implement an edge-detection procedure for images carried by the read beam. We set the reference beam to be a Gaussian beam, while the object beam carries an OAM beam with a topological charge of $\ell =+1$. As can be seen in the inset of Figure~1a, the interference structure of the reference and object beams induces a holographic metasurface with a fork grating in the ITO layer with a single dislocation. By diffracting an image carried by the read beam off such a fork grating corresponds to performing a generalized Hilbert transform or convolution tasks with special radially symmetric kernels \cite{furhapter2005spiral}. Since the transform at any point depends on the spatial structure of the whole grating, the transform is nonlocal in nature and can also be understood as a first-order optical differentiation operation \cite{zhu2021topological}. In fact, the order of the spatial differentiation can be controlled by the topological charge of the object beam. In our experiment, we input two different images (amplitude structures) encoded in the read beam (Fig.~\ref{fig.2d-differentiation}, left column). After undergoing a differentiation operation by the hologram, the resulting outputs are edge-detected images (Fig.~\ref{fig.2d-differentiation}, right column). The output images show a near-uniform circularly symmetric response of our optical system, highlighting the isotropic response ITO and the ability to process complex images. 
We note that although the response of the material is dominated by the hot electron effect, we have not observed any discernible degradation of the image quality during our experiments. This is due to short pulses that we have used and the slow nature of all relevant diffusive processes.
The example operations showcase the potential of our technique, which could be readily extended to more sophisticated holographic metastructures generated by multibeam interference and operations that could exploit both the image and conjugate image outputs. The versatility and rapid rewritability could enable complex computation tasks, including higher-order differentiation, integration, and equation solving \cite{cordaro2023solving, mohammadi2019inverse}.

\noindent
\textbf{Conclusions}\\
In summary, we have presented a nonlinear optical approach to efficient, rapidly reconfigurable real-time holographic protocols using a thin film of ITO. The demonstrated ENZ-based material platform is capable not only of direct information transfer through the spatial amplitude and phase distributions of beams but can also perform all-optical operations. The speed and efficiency with which this light-encoded information can be read, modified and re-written could make fast analog computation in scenarios where both time- and frequency-division multiplexed operation are required.

In this report, we have demonstrated proof-of-principle implementations of real-time holography using a nanomaterial. There are significant opportunities in improving the performance, including the energy expenditure, and functionality by orders of magnitude through judicious engineering. Given that the diffraction efficiency in our system is achieved by a combination of phase, absorption, and amplitude gratings -- due to simultaneous modification of the real and the imaginary parts of the refractive index-- we argue that manipulating the relative impact of each could boost the diffraction efficiency. Furthermore, the optimization of the linear optical properties and the dimensions of ENZ material, addition of plasmonic or dielectric structures, and implementation in reflection rather than transmission may also lead to increased efficiency while simultaneously decreasing the energy expenditure. Beyond the first-order differentiation operation shown here and the ready extension to higher-order differentiation, integration, and convolution operations, the nonlocal transformation could enable all-optical re-writable layers within computational neural networks, accelerate inverse design and prototyping systems, and facilitate adaptivity in multimode computation and imaging protocols. Consider, for example, an information-carrying beam experiencing spatio-temporal fluctuations from turbulence; a rapidly reconfigurable holographic surface could enable the recording and correction of disturbances through automatic phase conjugation in the diffraction process or through time gating. Finally, although the current scheme requires no nanofabrication, the diffraction efficiency could be enhanced while decreasing the intensity requirements, by orders of magnitude, by incorporating resonant nanostructured materials on the ITO layer\cite{alam2018large}. Beyond the boost in performance, inclusion of resonant plasmonic or silicon nanostructures could also allow polarization-dependent amplification, processing of information encoded in the spin-degree of freedom, and the transfer of information between spin and the angular momentum of light \cite{karimi2023interactions}. Lastly, future work may involve exploring the application of the presented method to recent approaches for temporal signal processing \cite{cotrufo2024temporal}, ultrafast modulation in space and time \cite{sisler2024electrically}, and reprogrammable real-time metasurface for machine learning and nonlinear optics~\cite{yanagimoto2025programmable,onodera2025arbitrary}.

\medskip
\textbf{Acknowledgements} \par 
We acknowledge support through the Natural Sciences and Engineering Research Council of Canada, the Canada Research Chairs program, and the Canada First Research Excellence Fund award on Transformative Quantum Technologies.  Additionally, RF acknowledges the support of the Research Council of Finland through the Photonics Research and Innovation Flagship (PREIN - decision 346511) and through the Academy Research Fellowship (decision 332399). In addition, RWB acknowledges support through the US Office of Naval Research MURI award N00014-20-1-2558, US National Science Foundation Award 2138174, and DOE award FWP 76295.

\medskip

%

\medskip
\textbf{Data availability} \par 
All underlying data are presented in the manuscript and the supplemental document. Raw data files may be obtained from the authors upon reasonable request.

\bibliography{bib-arxiv}

@article{marder1997design,
  title={Design and synthesis of chromophores and polymers for electro-optic and photorefractive applications},
  author={Marder, Seth R and Kippelen, Bernard and Jen, Alex K-Y and Peyghambarian, Nasser},
  journal={Nature},
  volume={388},
  number={6645},
  pages={845--851},
  year={1997},
  publisher={Nature Publishing Group UK London}
}

@article{blanche2010holographic,
  title={Holographic three-dimensional telepresence using large-area photorefractive polymer},
  author={Blanche, P-A and Bablumian, A and Voorakaranam, R and Christenson, C and Lin, W and Gu, T and Flores, D and Wang, P and Hsieh, W-Y and Kathaperumal, M and others},
  journal={Nature},
  volume={468},
  number={7320},
  pages={80--83},
  year={2010},
  publisher={Nature Publishing Group UK London}
}

@article{zhu2021topological,
  title={Topological optical differentiator},
  author={Zhu, Tengfeng and Guo, Cheng and Huang, Junyi and Wang, Haiwen and Orenstein, Meir and Ruan, Zhichao and Fan, Shanhui},
  journal={Nat. Commun.},
  volume={12},
  number={1},
  pages={680},
  year={2021},
  publisher={Nature Publishing Group UK London}
}

@article{mohammadi2019inverse,
  title={Inverse-designed metastructures that solve equations},
  author={Mohammadi Estakhri, Nasim and Edwards, Brian and Engheta, Nader},
  journal={Science},
  volume={363},
  number={6433},
  pages={1333--1338},
  year={2019},
  publisher={American Association for the Advancement of Science}
}

@article{silva2014performing,
  title={Performing mathematical operations with metamaterials},
  author={Silva, Alexandre and Monticone, Francesco and Castaldi, Giuseppe and Galdi, Vincenzo and Al{\`u}, Andrea and Engheta, Nader},
  journal={Science},
  volume={343},
  number={6167},
  pages={160--163},
  year={2014},
  publisher={American Association for the Advancement of Science}
}

@article{lopez2023self,
  title={Self-learning machines based on Hamiltonian echo backpropagation},
  author={Lopez-Pastor, Victor and Marquardt, Florian},
  journal={Phys. Rev. X},
  volume={13},
  number={3},
  pages={031020},
  year={2023},
  publisher={APS}
}

@article{Zheludev2012,
title={From metamaterials to metadevices},
author={Zheludev, Nikolay I and Kivshar, Yuri S},
journal={Nat. Mater.},
volume={11},
number={11},
pages={917--924},
year={2012},
publisher={Nature Publishing Group}
}

@article{Chen2016,
title={A review of metasurfaces: physics and applications},
author={Chen, Hou-Tong and Taylor, Antoinette J and Yu, Nanfang},
journal={Rep. Prog. Phys.},
volume={79},
number={7},
pages={076401},
year={2016},
publisher={IOP Publishing}
}

@article{LeCun2015,
title={Deep learning},
author={LeCun, Yann and Bengio, Yoshua and Hinton, Geoffrey},
journal={Nature},
volume={521},
number={7553},
pages={436--444},
year={2015},
publisher={Nature Publishing Group}
}

@article{Hesselink2004,
title={Holographic data storage},
author={Hesselink, Lambertus and Orlov, Sergei S and Bashaw, Matthew C},
journal={Proc. IEEE},
volume={92},
number={8},
pages={1231--1280},
year={2004},
publisher={IEEE}
}

@ARTICLE{Ashley2001,
  author={Ashley, J. and Bernal, M.-P and Burr, G. W. and Coufal, H. and Guenther, H. and Hoffnagle, J. A. and Jefferson, C. M. and Marcus, B. and Macfarlane, R. M. and Shelby, R. M. and Sincerbox, G. T.},
  journal={IBM J. Res. Dev.}, 
  title={Holographic data storage technology}, 
  year={2000},
  volume={44},
  number={3},
  pages={341-368},
}

@article{Barbastathis2019,
title={Intelligent imaging: Towards fully 3D optical sensing and reconstruction},
author={Barbastathis, George and Ozcan, Aydogan and Situ, Guohai},
journal={Opt. Express},
volume={27},
number={18},
pages={34323--34347},
year={2019},
publisher={Optical Society of America}
}

@article{Lin2017,
title={All-optical machine learning using diffractive deep neural networks},
author={Lin, Xing and Rivenson, Yair and Yardimci, Nezih T and Veli, Muhammed and Luo, Yi and Jarrahi, Mona and Ozcan, Aydogan},
journal={Science},
volume={361},
number={6406},
pages={1004--1008},
year={2018},
publisher={American Association for the Advancement of Science}
}

@article{Alam2016,
title={Large optical nonlinearity of indium tin oxide in its epsilon-near-zero region},
author={Alam, M Zahirul and De Leon, Israel and Boyd, Robert W},
journal={Science},
volume={352},
number={6287},
pages={795--797},
year={2016},
publisher={American Association for the Advancement of Science}
}

@article{Ozcan2016,
title={Mobile phones democratize and cultivate next-generation imaging, diagnostics and measurement tools},
author={Ozcan, Aydogan and McLeod, Euan},
journal={Annu. Rev. Biomed. Eng.},
volume={18},
pages={77--102},
year={2016},
publisher={Annual Reviews}
}

@article{allen1992orbital,
  title={Orbital angular momentum of light and the transformation of Laguerre-Gaussian laser modes},
  author={Allen, Les and Beijersbergen, Marco W and Spreeuw, R J C and Woerdman, J P},
  journal={Phys. Rev. A},
  volume={45},
  number={11},
  pages={8185},
  year={1992},
  publisher={APS}
}

@article{beijersbergen1993astigmatic,
  title={Astigmatic laser mode converters and transfer of orbital angular momentum},
  author={Beijersbergen, Marco W and Allen, Les and Van der Veen, HELO and Woerdman, JP},
  journal={Opt. Commun.},
  volume={96},
  number={1-3},
  pages={123--132},
  year={1993},
  publisher={Elsevier}
}

@article{vaity2012topological,
  title={Topological charge dependent propagation of optical vortices under quadratic phase transformation},
  author={Vaity, Pravin and Singh, RP},
  journal={Opt. Lett.},
  volume={37},
  number={8},
  pages={1301--1303},
  year={2012},
  publisher={Optica Publishing Group}
}

@article{furhapter2005spiral,
  title={Spiral phase contrast imaging in microscopy},
  author={F{\"u}rhapter, Severin and Jesacher, Alexander and Bernet, Stefan and Ritsch-Marte, Monika},
  journal={Opt. Express},
  volume={13},
  number={3},
  pages={689--694},
  year={2005},
  publisher={Optica Publishing Group}
}

@article{alam2018large,
  title={Large optical nonlinearity of nanoantennas coupled to an epsilon-near-zero material},
  author={Alam, M. Zahirul and Schulz, Sebastian A and Upham, Jeremy and De Leon, Israel and Boyd, Robert W},
  journal={Nat. Photonics},
  volume={12},
  number={2},
  pages={79--83},
  year={2018},
  publisher={Nature Publishing Group UK London}
}

@article{kulce2021all,
  title={All-optical synthesis of an arbitrary linear transformation using diffractive surfaces},
  author={Kulce, Onur and Mengu, Deniz and Rivenson, Yair and Ozcan, Aydogan},
  journal={Light: Sci. Appl.},
  volume={10},
  number={1},
  pages={196},
  year={2021},
  publisher={Nature Publishing Group UK London}
}

@article{gigan2022imaging,
  title={Imaging and computing with disorder},
  author={Gigan, Sylvain},
  journal={Nat. Phys.},
  volume={18},
  number={9},
  pages={980--985},
  year={2022},
  publisher={Nature Publishing Group UK London}
}

@article{wang2023image,
  title={Image sensing with multilayer nonlinear optical neural networks},
  author={Wang, Tianyu and Sohoni, Mandar M and Wright, Logan G and Stein, Martin M and Ma, Shi-Yuan and Onodera, Tatsuhiro and Anderson, Maxwell G and McMahon, Peter L},
  journal={Nat. Photonics},
  volume={17},
  number={5},
  pages={408--415},
  year={2023},
  publisher={Nature Publishing Group UK London}
}

@article{gabor1971holography,
  title={Holography: The fundamentals, properties, and applications of holograms are reviewed.},
  author={Gabor, Dennis and Kock, Winston E and Stroke, George W},
  journal={Science},
  volume={173},
  number={3991},
  pages={11--23},
  year={1971},
  publisher={American Association for the Advancement of Science}
}

@article{goodman1971introduction,
  title={An introduction to the principles and applications of holography},
  author={Goodman, Joseph W},
  journal={Proc. IEEE},
  volume={59},
  number={9},
  pages={1292--1304},
  year={1971},
  publisher={IEEE}
}

@article{stroke1972optical,
  title={Optical computing},
  author={Stroke, George W},
  journal={IEEE spectrum},
  volume={9},
  number={12},
  pages={24--41},
  year={1972},
  publisher={IEEE}
}

@article{volodin1995highly,
  title={Highly efficient photorefractive polymers for dynamic holography},
  author={Volodin, Boris L and Meerholz, Klaus and Kippelen, Bernard and Kukhtarev, Nickolai V and Peyghambarian, Nasser and others},
  journal={Opt. Eng.},
  volume={34},
  number={8},
  pages={2213--2223},
  year={1995},
  publisher={SPIE}
}

@article{miller2017attojoule,
  title={Attojoule optoelectronics for low-energy information processing and communications},
  author={Miller, David A B},
  journal={J. Lightwave Technol.},
  volume={35},
  number={3},
  pages={346--396},
  year={2017},
  publisher={IEEE}
}

@article{zhang2021reconfigurable,
  title={Reconfigurable metasurface for image processing},
  author={Zhang, Xiaomeng and Zhou, You and Zheng, Hanyu and Linares, Alberto Esteban and Ugwu, Fabian Chibuzor and Li, Deyu and Sun, Hong-Bo and Bai, Benfeng and Valentine, Jason G},
  journal={Nano Lett.},
  volume={21},
  number={20},
  pages={8715--8722},
  year={2021},
  publisher={ACS Publications}
}

@article{wetzstein2020inference,
  title={Inference in artificial intelligence with deep optics and photonics},
  author={Wetzstein, Gordon and Ozcan, Aydogan and Gigan, Sylvain and Fan, Shanhui and Englund, Dirk and Solja{\v{c}}i{\'c}, Marin and Denz, Cornelia and Miller, David AB and Psaltis, Demetri},
  journal={Nature},
  volume={588},
  number={7836},
  pages={39--47},
  year={2020},
  publisher={Nature Publishing Group UK London}
}

@article{ambs2010optical,
  title={Optical computing: a 60-year adventure},
  author={Ambs, Pierre},
  journal={Adv. Opt. Technol.},
  volume={2010},
  number={1},
  pages={372652},
  year={2010},
  publisher={Wiley Online Library}
}

@incollection{goodman2020short,
  title={A short history of the field of optical computing},
  author={Goodman, Joseph W},
  booktitle={Optical Computing},
  pages={7--21},
  year={2020},
  publisher={CRC Press},
    address={Boca Raton}
}

@article{solli2015analog,
  title={Analog optical computing},
  author={Solli, Daniel R and Jalali, Bahram},
  journal={Nat. Photonics},
  volume={9},
  number={11},
  pages={704--706},
  year={2015},
  publisher={Nature Publishing Group UK London}
}

@article{hu2024diffractive,
  title={Diffractive optical computing in free space},
  author={Hu, Jingtian and Mengu, Deniz and Tzarouchis, Dimitrios C and Edwards, Brian and Engheta, Nader and Ozcan, Aydogan},
  journal={Nat. Commun.},
  volume={15},
  number={1},
  pages={1525},
  year={2024},
  publisher={Nature Publishing Group UK London}
}

@book{goodman2005fourieroptics,
  title={Introduction to Fourier optics},
  author={Goodman, Joseph W},
  year={2005},
  publisher={Roberts and Company publishers},
    address={Greenwood Village, CO}
}

@article{shen2019optical,
  title={Optical vortices 30 years on: OAM manipulation from topological charge to multiple singularities},
  author={Shen, Yijie and Wang, Xuejiao and Xie, Zhenwei and Min, Changjun and Fu, Xing and Liu, Qiang and Gong, Mali and Yuan, Xiaocong},
  journal={Light: Sci. Appl.},
  volume={8},
  number={1},
  pages={90},
  year={2019},
  publisher={Nature Publishing Group UK London}
}

@article{yao2011orbital,
  title={Orbital angular momentum: origins, behavior and applications},
  author={Yao, Alison M and Padgett, Miles J},
  journal={Adv. Opt. Photonics},
  volume={3},
  number={2},
  pages={161--204},
  year={2011},
  publisher={Optica Publishing Group}
}

@article{reshef2019nonlinear,
  title={Nonlinear optical effects in epsilon-near-zero media},
  author={Reshef, Orad and De Leon, Israel and Alam, M Zahirul and Boyd, Robert W},
  journal={Nat. Rev. Mater.},
  volume={4},
  number={8},
  pages={535--551},
  year={2019},
  publisher={Nature Publishing Group UK London}
}

@article{kinsey2019near,
  title={Near-zero-index materials for photonics},
  author={Kinsey, Nathaniel and DeVault, Clayton and Boltasseva, Alexandra and Shalaev, Vladimir M},
  journal={Nat. Rev. Mater.},
  volume={4},
  number={12},
  pages={742--760},
  year={2019},
  publisher={Nature Publishing Group UK London}
}

@article{mecher2002near,
  title={Near-infrared sensitivity enhancement of photorefractive polymer composites by pre-illumination},
  author={Mecher, Erwin and Gallego-G{\'o}mez, Francisco and Tillmann, Hartwig and H{\"o}rhold, Hans-Heinrich and Hummelen, Jan C and Meerholz, Klaus},
  journal={Nature},
  volume={418},
  number={6901},
  pages={959--964},
  year={2002},
  publisher={Nature Publishing Group UK London}
}

@article{tsutsumi2016molecular,
  title={Molecular design of photorefractive polymers},
  author={Tsutsumi, Naoto},
  journal={Polym. J.},
  volume={48},
  number={5},
  pages={571--588},
  year={2016},
  publisher={Nature Publishing Group}
}

@article{cordaro2023solving,
  title={Solving integral equations in free space with inverse-designed ultrathin optical metagratings},
  author={Cordaro, Andrea and Edwards, Brian and Nikkhah, Vahid and Al{\`u}, Andrea and Engheta, Nader and Polman, Albert},
  journal={Nat. Nanotechnol.},
  volume={18},
  number={4},
  pages={365--372},
  year={2023},
  publisher={Nature Publishing Group UK London}
}

@article{karimi2023interactions,
  title={Interactions of Fundamental Mie Modes with Thin Epsilon-near-Zero Substrates},
  author={Karimi, Mohammad and Awan, Kashif Masud and Vaddi, Yaswant and Alaee, Rasoul and Upham, Jeremy and Alam, M. Zahirul and Boyd, Robert W},
  journal={Nano Lett.},
  volume={23},
  number={24},
  pages={11555--11561},
  year={2023},
  publisher={ACS Publications}
}

@article{yariv1978four,
  title={Four wave nonlinear optical mixing as real time holography},
  author={Yariv, Amnon},
  journal={Opt. Commun.},
  volume={25},
  number={1},
  pages={23--25},
  year={1978},
  publisher={Elsevier}
}

@article{magnusson1978diffraction,
  title={Diffraction regimes of transmission gratings},
  author={Magnusson, R and Gaylord, T K},
  journal={J. Opt. Soc. Am.},
  volume={68},
  number={6},
  pages={809--814},
  year={1978},
  publisher={Optica Publishing Group}
}

@article{ragnarsson1970a_new_holographic,
year = {1970},
month = {oct},
publisher = {},
volume = {2},
number = {4-5},
pages = {145},
author = {S. I. Ragnarsson},
title = {A new Holographic Method of Generating a High Efficiency, Extended Range Spatial Filter with Application to Restoration of Defocussed Images},
journal = {Phys. Scr.}
}

@article{tay_updatable_2008,
	title = {An updatable holographic three-dimensional display},
	volume = {451},
	pages = {694--698},
	number = {7179},
	journal = {Nature},
    year = {2008},
	author = {Tay, S. and Blanche, P.-A. and Voorakaranam, R. and Tun\c{c}, A. V. and Lin, W. and Rokutanda, S. and Gu, T. and Flores, D. and Wang, P. and Li, G. and St Hilaire, P. and Thomas, J. and Norwood, R. A. and Yamamoto, M. and Peyghambarian, N.},
}

@article{liebel2021ultrafast,
  title={Ultrafast transient holographic microscopy},
  author={Liebel, Matz and Camargo, Franco V A and Cerullo, Giulio and van Hulst, Niek F},
  journal={Nano Lett.},
  volume={21},
  number={4},
  pages={1666--1671},
  year={2021},
  publisher={ACS Publications}
}

@article{zhou2024optical,
  title={Optical computing metasurfaces: applications and advances},
  author={Zhou, Hongqiang and Zhao, Chongli and He, Cong and Huang, Lingling and Man, Tianlong and Wan, Yuhong},
  journal={Nanophotonics},
  volume={13},
  number={4},
  pages={419--441},
  year={2024},
  publisher={De Gruyter}
}

@article{kuznetsov2024roadmap,
  title={Roadmap for optical metasurfaces},
  author={Kuznetsov, Arseniy I and Brongersma, Mark L and Yao, Jin and Chen, Mu Ku and Levy, Uriel and Tsai, Din Ping and Zheludev, Nikolay I and Faraon, Andrei and Arbabi, Amir and Yu, Nanfang and others},
  journal={ACS Photonics},
  volume={11},
  number={3},
  pages={816--865},
  year={2024},
  publisher={ACS Publications}
}

@article{cotrufo2024reconfigurable,
  title={Reconfigurable image processing metasurfaces with phase-change materials},
  author={Cotrufo, Michele and Sulejman, Shaban B and Wesemann, Lukas and Rahman, Md Ataur and Bhaskaran, Madhu and Roberts, Ann and Al{\`u}, Andrea},
  journal={Nat. Commun.},
  volume={15},
  number={1},
  pages={4483},
  year={2024},
  publisher={Nature Publishing Group UK London}
}

@article{abou2025programmable,
  title={Programmable metasurfaces for future photonic artificial intelligence},
  author={Abou-Hamdan, Loubnan and Marinov, Emil and Wiecha, Peter and del Hougne, Philipp and Wang, Tianyu and Genevet, Patrice},
  journal={Nat. Rev. Phys.},
  pages={1--17},
  year={2025},
  publisher={Nature Publishing Group UK London}
}

@article{sisler2024electrically,
  title={Electrically tunable space--time metasurfaces at optical frequencies},
  author={Sisler, Jared and Thureja, Prachi and Grajower, Meir Y and Sokhoyan, Ruzan and Huang, Ivy and Atwater, Harry A},
  journal={Nat. Nanotechnol.},
  volume={19},
  number={10},
  pages={1491--1498},
  year={2024},
  publisher={Nature Publishing Group UK London}
}

@article{cotrufo2024temporal,
  title={Temporal signal processing with nonlocal optical metasurfaces},
  author={Cotrufo, Michele and Esfahani, Sedigheh and Korobkin, Dmitriy and Al{\`u}, Andrea},
  journal={npj Nanophotonics},
  volume={1},
  number={1},
  pages={39},
  year={2024},
  publisher={Nature Publishing Group UK London}
}

@article{onodera2025arbitrary,
  title={Arbitrary control over multimode wave propagation for machine learning},
  author={Onodera, Tatsuhiro and Stein, Martin M and Ash, Benjamin A and Sohoni, Mandar M and Bosch, Melissa and Yanagimoto, Ryotatsu and Jankowski, Marc and McKenna, Timothy P and Wang, Tianyu and Shvets, Gennady and others},
  journal={Nature Physics},
  pages={1--8},
  year={2025},
  publisher={Nature Publishing Group UK London}
}

@article{yanagimoto2025programmable,
  title={Programmable on-chip nonlinear photonics},
  author={Yanagimoto, Ryotatsu and Ash, Benjamin A and Sohoni, Mandar M and Stein, Martin M and Zhao, Yiqi and Presutti, Federico and Jankowski, Marc and Wright, Logan G and Onodera, Tatsuhiro and McMahon, Peter L},
  journal={Nature},
  pages={1--8},
  year={2025},
  publisher={Nature Publishing Group UK London}
}

\pagebreak
\onecolumngrid

\section{Supplemental document}

\subsection{Material Properties}

\begin{figure}[ht]
\centering
\includegraphics[width=.7\columnwidth]{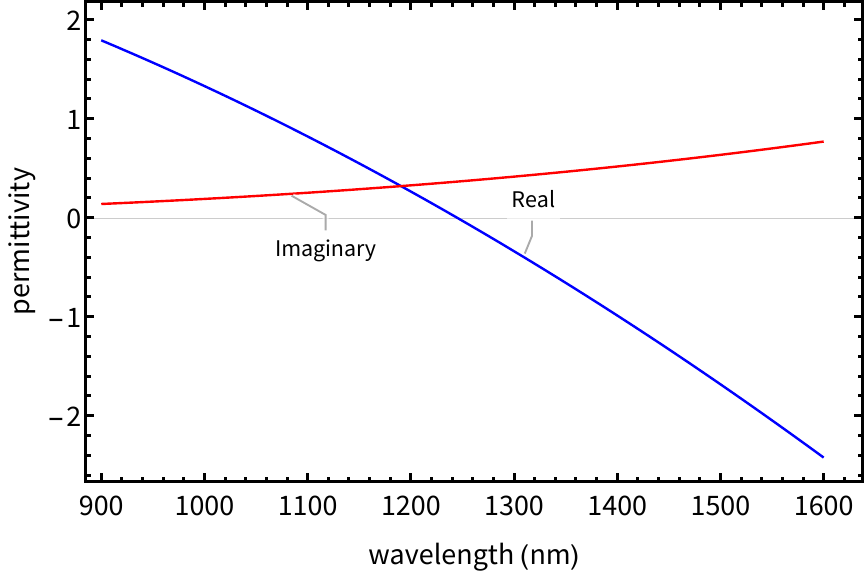}
\caption{\label{fig.permnittivity}
Permittivity of the ITO film used.}
\end{figure}

We purchased ITO films deposited on float glass substrates from Precision Glass \& Optics (PG\&O). We present the permittivity data of the material as obtained using ellipsometry in Fig.~\ref{fig.permnittivity}. The zero permittivity wavelength of the film is 1240~nm. The Drude parameters we used to model the permittivity of the material near the zero permittivity wavelength are $\epsilon_{\infty} = 3.77$, plasma frequency $\omega_p=2.954\times10^{15}$~rad/s and damping coefficient $\Gamma=0.0495\omega_p$.

\subsection{Experimental Setup}

\begin{figure}[ht]
\centering
\includegraphics[width=.9\columnwidth]{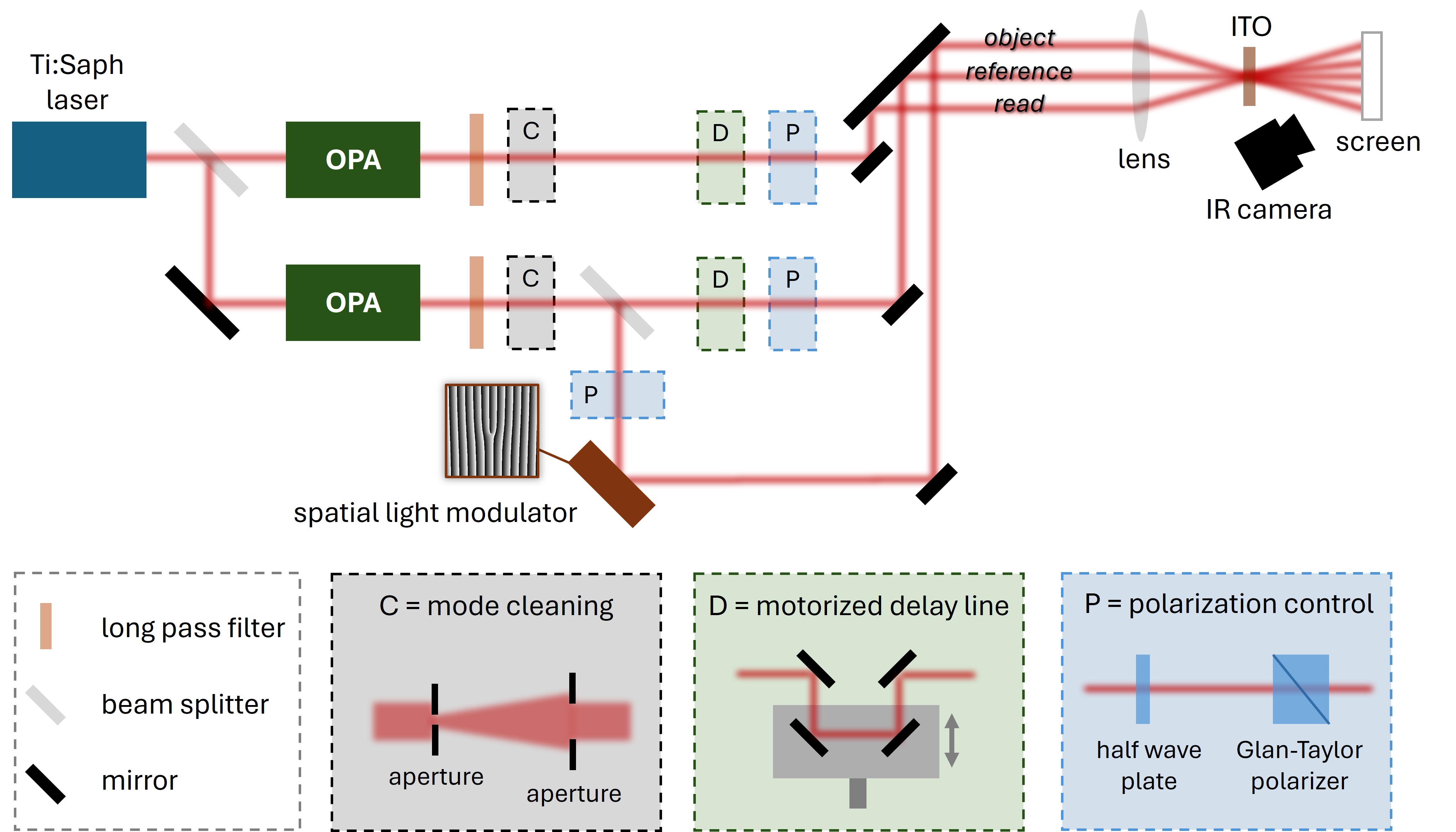}
\caption{\label{fig.setup}
Simplified sketch of the experimental setup.
}
\end{figure}

We use a pair of optical parametric amplifiers (OPA) pumped by a modelocked Ti:Saph laser producing 120-fs pulses at a wavelength of approximately 1260\,nm and a repetition rate of 1\,kHz. For both outputs, we remove residual light of a wavelength shorter than 1000\,nm using a long-pass filter, and we clean the spatial mode through a pair of apertures. We use the beam of one OPA as the read beam such that we can adjust its wavelength separately. We split the beam of the other OPA using a thin-film beam splitter to obtain the reference and the object beams, see Fig.~\ref{fig.setup} . Using a half-wave plate followed by a Glan-Taylor polarizer in each of the three beams, we ensure that all beams are p-polarized and their respective intensities are independently controllable.

Additionally, we control the length of the beam paths of the read and the reference beam using a pair of motorized delay lines, such that we can ensure the simultaneous arrival of all three pulses on the ITO sample, and we can also perform the pump-probe experiment with controllable time delay as described in the main text. The object beam is structured using computer-generated holograms implemented by a liquid-crystal phase-only spatial light modulator. 
We use a Hamamatsu spatial light modulator displaying computer generated holograms to generate the spatial modes (see inset for an example of a hologram imprinting an OAM mode of $\ell=1$). Although the SLM is not designed to be used at wavelengths of interest for this work, the SLM  is functional at these wavelengths but with a a slightly reduced diffraction efficiency. We used Matlab to generate the holograms for the spatial modes. The letter ``F'' was made by cutting a thin anodized aluminum sheet. The maple leaf structure was a metal mask made using a commercial laser cutting service. Finally, we focus all three beams jointly onto the ITO using a single two-inch lens and record all beams, including the newly generated image beams, using a white screen, which we image by an IR camera. For all efficiency measurements given in the main text, we used the screen and camera only to simplify the alignment but recorded all power values by putting a power meter in the required beam path (not shown in Fig.~\ref{fig.setup}).

\subsection{Unprocessed camera recording}

\begin{figure}[htb]
\centering
\includegraphics[width=.99\columnwidth]{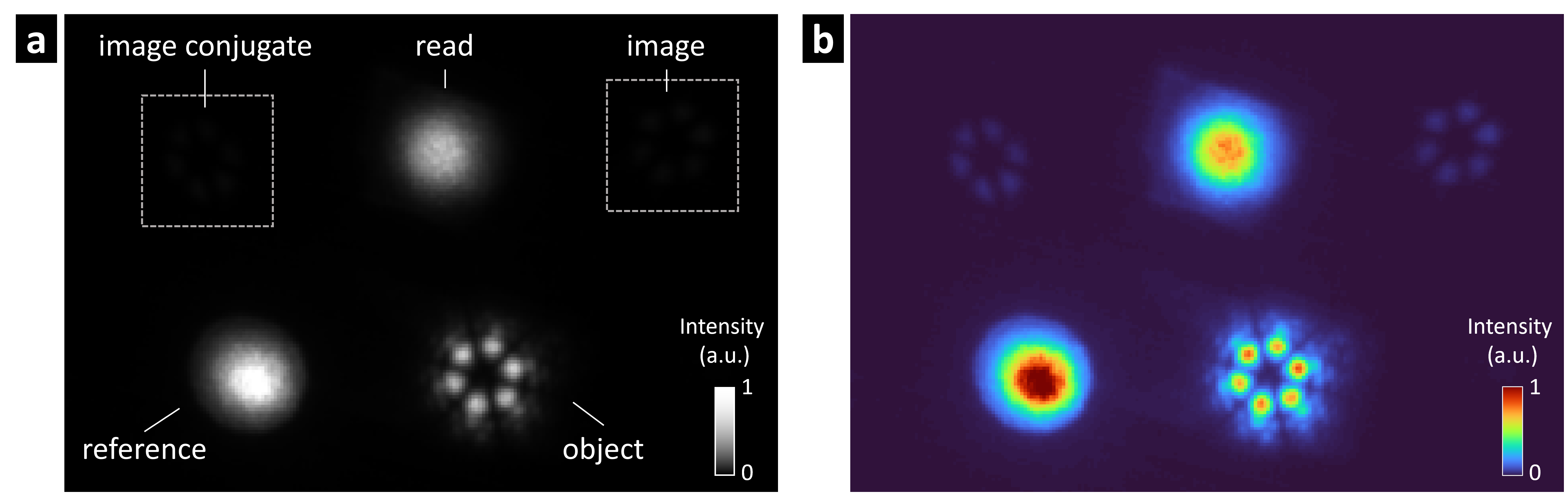}
\caption{\label{fig.original_camera}
Original recording of all beams involved in the experiment.
a) The unprocessed image shows all beams, object, reference, read, image, and image conjugate as labeled in the main text. However, the image and image conjugate are barely visible because of the low dynamic range of the camera. b) Using a different colormap (see inset), both image beams become better visible, without the need to brighten the regions.
}
\end{figure}

In Fig.~\ref{fig.original_camera}(a) we present the unprocessed version of the camera recording shown in Fig.\,1 in the main text. In the main text we adjust two regions where the two beams, image and conjugate beam, are brightened by a factor of 10 for better visibility. Using a different colormap, the image and image conjugate beams become better visible without the need to increase the brightness of the signal and conjugate beams, see Fig.~\ref{fig.original_camera}(b).

\end{document}